

\input amstex
\documentstyle{amsppt}
\nologo
\pagewidth{160truemm}
\pageheight{240truemm}

\newwrite\auxout
\newif\ifauxopen
\newif\iffileexists
\chardef\letter=11
\def\setletters{%
 \catcode`\:=\letter \catcode`\;=\letter \catcode`\.=\letter
 \catcode`\0=\letter \catcode`\1=\letter \catcode`\2=\letter
 \catcode`\3=\letter \catcode`\4=\letter \catcode`\5=\letter
 \catcode`\6=\letter \catcode`\7=\letter \catcode`\8=\letter
 \catcode`\9=\letter}
\def\crossrefs{%
 \begingroup
 \immediate\openin0=\jobname.aux\space
 \ifeof0 \global\fileexistsfalse
  \message{Creating file \jobname.aux}
  \else\global\fileexiststrue\fi
 \immediate\closein0\endgroup
 \iffileexists\begingroup\setletters
   \input \jobname.aux \endgroup \fi
 \immediate\openout\auxout=\jobname.aux
   \message{Writing to \jobname.aux}
   \auxopentrue}%
\def\endcrossrefs{\ifauxopen\immediate\closeout\auxout\fi}
\def\xlabel#1{\ifauxopen \begingroup\setletters 
 \expandafter\xdef\csname XR#1\endcsname{\xreflabel}%
 \immediate\write\auxout{\string\gdef
   \expandafter\string\csname XR#1\endcsname{\xreflabel}}%
\endgroup\ignorespaces\fi}%
\def\setlabel#1{\global\let\xreflabel=#1}
\def\stepsetlabel#1#2{\stepup{#1}\setlabel{#2}#2}
\setlabel\null
\def\xref#1{\begingroup\setletters
 \expandafter\ifx \csname XR#1\endcsname \relax
   \message{Label `#1' undefined} %
    \expandafter\gdef \csname XR#1\endcsname{00}%
  \else \csname XR#1\endcsname
\fi\endgroup}

\def\stepup#1{\global\advance#1by1}
\def\stepdown#1{\global\advance#1by-1}
\newcount\eqct
\newcount\secct
\newcount\procct
\newcount\defct
\def\eqlb{\number\eqct}
\def\seclb{\number\secct}
\def\proclb{\number\procct}
\def\deflb{\number\defct}
\def\eqnum{(\stepsetlabel\eqct\eqlb)}
\def\secnum{\stepsetlabel\secct\seclb}
\def\procnum{\stepsetlabel\procct\proclb}
\def\defnum{\stepsetlabel\defct\deflb}
%

%
%
\def\du#1#2{{}_{#1}{}^{#2}}       
\def\ud#1#2{{}^{#1}{}_{#2}}       
\def\st#1#2{{}^{#1}_{#2}}         
\def\gt#1#2#3#4{\st{#1\ldots#2}{#3\ldots#4}}
%
%
\def\tCD{{\tilde{\cal D}}}
\def\CC{{\cal C}}               
\def\cD{{\cal D}}               
\def\CA#1{\tilde{\cal A}_{#1}}  
\def\OCA#1{{\cal A}_{#1}}       
\def\CT{{\cal T}}               
\def\CE{{\cal E}}               
\def\CM{{\cal E}_M}             
\def\CN{{\cal N}}               
\def\CG{{\cal G}}               
\def\CK{{\cal K}}               
%
%
\def\mmu{{\mu^{-1}}}            
\def\mux#1#2{\mu^{#1}_{#2}}     
\def\mmux#1#2{\eta^{#1}_{#2}}   
\def\mujac{J_\mu}               
\def\mufun{\mu^*}               
\def\muphi{\hat\mu_*}           
\def\muem{\tilde\mu^*}          
%
%
\def\Phe{\Phi_\eps}
\def\Phicov{\Phi\du}     
\def\Phivec{\Phi\ud}     
\def\Pex{\Phi^{(\eps)}_x}  
\def\Pexcov{\Pex\du}     
\def\Pexvec{\Pex\ud}     
\def\Pxe{\Phi_{x,\eps}}    
%
%
\def\transp{\Gamma}                          
\def\trcov{\Gamma\du}   
\def\trvec{\Gamma\ud}     
\def\trcovp{\Gamma'\du}   
\def\con{\gamma\st}
\def\conp{\gamma'\st}
\def\Con{\Gamma\st}
\def\hcon{\hat\gamma\st}
\def\hCon{\hat\Gamma\st}
%
%
\def\tGamma{{\tilde\Gamma}}
\def\tomega{{\tilde\omega}}
\def\tf{{\tilde f}}
\def\tS{{\tilde S}}
\def\tT{{\tilde T}}
\def\tX{{\tilde X}}
\def\tR{{\tilde R}}
\def\tg{{\tilde g}}
%
%
\def\pbyp#1#2{{\partial#1\over\partial#2}}
\def\pbypmix#1#2#3{{\partial^2 #1 \over \partial #2 \partial #3}}
%
%
\def\kron{\delta\st}              
\def\Diff#1#2{\hbox{\rm Diff}(#1,#2)}  
\def\compact{\subset\subset}      
\def\halfk{{\lfloor k/2 \rfloor}} 
\def\Real{{\Bbb R}}               
\def\Nat{{\Bbb N}}                
\def\tang{{\Bbb T}}
\def\cotang{{\Bbb T}^*}
\let\eps=\varepsilon              
\let\leq=\leqslant                

\let\geq=\geqslant

\def\tprod#1#2{\left\langle #1,#2 \right\rangle}  
\def\implies{\;\Longrightarrow\;} 
\def\limeps{\lim_{\eps\to0}}      
\let\embed=\iota                  
\def\Lie{{\cal L}}                
\def\eqref#1{(\xref{#1})}
\let\cal=\Cal
\def\emph#1{{\it#1}}


\topmatter
\title
A nonlinear theory of tensor distributions
\endtitle
\author
J.A. Vickers and J.P. Wilson
\endauthor
\address
Faculty of Mathematical Studies,
University of Southampton,
Southampton  SO17 1BJ, UK.
\endaddress
\email\nofrills
\emph{Email Addresses}:
jav\@maths.soton.ac.uk,
jpw\@maths.soton.ac.uk
\endemail
\subjclass\nofrills
PACS: 0240, 0420
\endsubjclass
\abstract
The coordinate invariant theory of generalised functions of Colombeau and
Meril is reviewed and extended to enable the construction of multi-index
generalised tensor functions whose transformation laws coincide with their
counterparts in classical distribution theory.
\endabstract
\endtopmatter

\document
\crossrefs

\head
\secnum.
Introduction
\endhead
Colombeau's theory of new generalised functions (Colombeau, 1984) has
increasingly had an important role to play in General Relativity, enabling
a distributional interpretation to be given to products of distributions
which would otherwise be undefined in the framework of Classical
distribution theory. Recent applications of Colombeau's theory to
Relativity have included the calculation of distributional curvatures which
correspond to metrics of low differentiability, such as those which occur
in space-times with thin cosmic strings (Clarke et al, 1996; Wilson 1997)
and Kerr singularities (Balasin, 1997), and the electromagnetic field
tensor of ultra-relativistic Riessner-Nordstr{\o}m solution (Steinbauer,
1997).

The theory of General Relativity is built upon the Principle of General
Covariance which means that the measurement of all physical quantities do
not depend on what local coordinate system the observer is using; therefore
they must transform locally as tensors under $C^\infty$ diffeomorphisms.
Unfortunately, Colombeau's theory is not naturally suited for representing
such a covariant physical theory because it is heavily built upon the
linear structure.  Colombeau's algebra $\CG(\Omega)$ is a quotient algebra
on the algebra $\CM(\Omega)$ of smooth functionals
$\OCA{0}\times\Omega\to\Real$, where one defines the kernel spaces
$\OCA{k}$ as being the subspaces of functions $\Phi\in\cD$ such that
$$\eqalignno{
&\int \Phi(\xi) d\xi=1  &\eqnum\xlabel{mom1} \cr
&\int \xi^\alpha \Phi(\xi) \,d\xi =
   0 \qquad \alpha\in\Nat^n, \quad 1\leq|\alpha|\leq k, 
&
\eqnum\xlabel{mom2}
\cr}
$$
These spaces are not invariant under a smooth diffeomorphism $x'=\mu(x)$,
unless $\mu$ is linear. Since these spaces are used in both the definitions
of the spaces $\CM(\Omega)$ (moderate functions) and $\CN(\Omega)$ (null
functions) it makes the definition of a mapping $\muem:\CE(\Omega') \to
\CE(\Omega)$ which preserves both $\CM(\Omega)$ and $\CN(\Omega)$ very
problematic. 

As a consequence of this, one must apply Colombeau's theory with great
caution in that coordinate transformations should be carried out at the
distributional level and Colombeau's algebra is to be used as a tool for
assigning a distributional interpretation to nonlinear operations involving
distributions. It was shown by Vickers and Wilson (1998) that the
calculation of Clarke et al.\ (1996), which shows that the distributional
Ricci scalar density of a conical singularity with a deficit angle of
$2\pi(1-A)$ is $4\pi(1-A)\delta^{(2)}$, is invariant under $C^\infty$
transformations in the sense that the following diagram commutes
$$
\CD
   {g'_{ab}} @>{\embed'}>> {\widetilde{g'_{ab}}} @>>>
   {\tR'\sqrt{\tg'}} @>{\approx}>> {4\pi(1-A)\delta^{(2)}} \\
   @VV{\mu^*}V &&&&  @VV{\mu^*}V \\
   {g_{ab}} @>{\embed}>> {\widetilde{g_{ab}}} @>>>
   {\tR\sqrt{\tg}} @>{\approx}>> {4\pi(1-A)\delta^{(2)}}
\endCD
$$   
where the maps $\mu^*$ denote the relevant tensor distribution
transformations. It should be noted that Colombeau's theory is applied to
both coordinate systems and that no attempt is made to compare generalised
functions in one coordinate system with their counterparts until the
distributional curvature has been recovered via weak equivalence.

The problem resulting from the inability of the spaces $\OCA{k}$ to
transform invariantly has been overcome by the use of the `simplified'
algebra $\CG_s(\Omega)$ which is formed from using a base algebra
$\CE_s(\Omega)$ consisting of smooth functionals on $(0,1]\times\Omega$
rather than on $\OCA{k}\times\Omega$ (Biagioni, 1990; Colombeau
1992), so one does not need to worry about how $\OCA{k}(\Omega)$ transforms
under $\mu$, and hence is able to define a natural map
$\muem:\CE_s(\Omega')\to\CE_s(\Omega)$ which preserves its moderate
subalgebra $\CE_{M,s}(\Omega')$ and its null ideal $\CN_s(\Omega')$.  However
there is still a problem which undermines the suitability of using
$\CG(\Omega)$ as a coordinate invariant representation of distribution
theory, which itself is a coordinate invariant construction, in that the
embedding $\embed:\cD'(\Omega)\to\CM(\Omega)$ provided by the smoothing
convolution integral does not commute with the diffeomorphism $\mu$. 

A  solution to this problem was proposed by Colombeau and
Meril (1994), in which they constructed a coordinate invariant
$\CG(\Omega)$ together with a scalar transformation that commutes with the
embedding. It essentially involves weakening the moment conditions used in
defining the smoothing kernel space $\OCA{k}$ which is then defined
in a coordinate invariant manner. 


In this paper we shall begin by reviewing the algebra of Colombeau and
Meril and extend the definitions so that multi-index tensor transformations
which commute with the embedding may also be defined.

\head
\secnum.
Scalars
\endhead
Let $\Omega$ and $\Omega'$ be open subsets of $\Real^n$ that are
diffeomorphic to each other by $\mu:\Omega\to\Omega'$.  In general the map
$\mu$ will induce a map $\mufun:\CC(\Omega')\to\CC(\Omega)$, between the
spaces of continuous functions on $\Omega$ and $omega$, given by
$\mufun(f')=f'\circ\mu$, enabling $f'$ to transform as a scalar between the
two coordinate systems.

Functions from $\CC(\Omega)$ may be embedded into Colombeau's generalised
function algebra $\CG(\Omega)$ by the smoothing convolution
$$ \eqalign{ & \embed: \CC(\Omega) \to \CM(\Omega) \cr & \embed(f) = \tf  \cr }$$
where
$$ \tf(\Phi,x)=\int f(x+\xi) \Phi(\xi) \,d\xi \qquad \Phi\in\OCA{k}
$$ 
We would like to be able to construct a map $\muem:\CM(\Omega')\to\CM(\Omega)$,
an analogue of $\mufun$, such that
$$
\CD
   \CC(\Omega') @>{\embed'}>> \CM(\Omega') \cr
   @VV{\mufun}V @VV{\muem}V \cr
   \CC(\Omega) @>{\embed}>>  \CM(\Omega) \cr
\endCD
$$
commutes, and furthermore we would like $\muem$ to map $\CN(\Omega')$ to
$\CN(\Omega)$ so that we can construct $\CG(\Omega')$ and $\CG(\Omega)$
consistently.

Given $ \tf' \in \CM(\Omega') $, the obvious way to construct $\muem\tf'$ is
by defining
$$ \muem\tf(\Phi,x)=\tf(\muphi\Phi,\mu(x)) $$
where $\muphi$ is some mapping on the kernel space $\OCA{k}$.  If we
take $\muphi(\Phi)=\Phi\circ\mmu$ then $\muphi(\Phi)$ will no longer
satisfy the normalisation condition
$$ \int \muphi\Phi(\lambda) \,d\lambda =1 $$
However if we take
$$\muphi(\Phi) = {1\over|\mujac|} \Phi\circ\mmu$$
then the normalisation is preserved. With the choice above,
$$ \eqalign{
   \muem\tf'(\Phi,x) &= \int f'(\mu(x)+\lambda)
   {1\over \mujac(\mmu(\lambda))} \Phi(\mmu(\lambda)) \,d\lambda \cr
   &=\int f'(\mu(x)+\mu(\xi)) \Phi(\xi) \, d\xi \cr
}$$
On the other hand
$$ \widetilde{f'\circ\mu} (\Phi,x) =\int f'(\mu(x+\xi))
   \Phi(\xi) \, d\xi $$
so  $\muem\tf'$ and $\widetilde{f'\circ\mu}$ are different.

In order for $\muem\tf'$ and $\widetilde{f'\circ\mu}$ to represent the same
generalised function, their difference must be null; however, the problem
is that there is no obvious relation between $\muphi(\OCA{k})$ and
$\OCA{p}$. This prevents us from having a coordinate invariant
definition in the framework of Colombeau's original theory.

This problem was resolved by Colombeau and Meril~(1994) in which they
gave a new definition of the space of smoothing kernels, $\CA{k}$ together with
a map $\muphi$ in such a way that
$\muem\tf'=\widetilde{f'\circ\mu}$ and the notion of a null function was
preserved.

It is convenient to start by considering generalised functions on
$\Real^n$ together with diffeomorphisms $\mu:\Real^n \to \Real^n$. We
will later restrict the algebra to open sets $\Omega$ and consider
diffeomorphisms between open sets. We begin by considering the
embedding of $f \circ \mu$ using the $\eps$-dependent kernel
$\Phi_\eps(x)={1 \over {\eps^n}}\Phi\left({x \over \eps}\right)$
$$\eqalign{
\widetilde{f'\circ\mu} (\Phi_\eps,x) &={1 \over {\eps^n}}
\int f'(\mu(x+\xi))\Phi(\xi/\eps) \, d\xi \cr 
&=\int f'(\mu(x+\eps\eta))\Phi(\eta) \, d\eta \cr
&=\int {f'(\mu(x)+\eps\lambda)\over\left|\mujac(\mmu(\mu(x)+\eps\lambda)) 
\right|}  \Phi\left({\mmu(\mu(x)+\eps\lambda)
-x\over\eps}\right)\,d\lambda \cr}
$$
where $\lambda$ is defined by
$\mu(x+\eps\eta)=\mu(x)+\eps\lambda$.

If we now \emph{define}
$$
\muphi\Phi(\lambda)={1\over\left|\mujac(\mmu(\mu(x)+\eps\lambda))
   \right|} \Phi \left({\mmu(\mu(x)+\eps\lambda)-x\over\eps}\right)
$$
then
$$\eqalign{
\widetilde{f'\circ\mu} (\Phi_\eps,x) 
&=\int f'(\mu(x)+\eps\lambda))\left(\muphi\Phi(\lambda)\right)
\, d\lambda \cr
&={1 \over {\eps^n}}
\int f'(\mu(x)+\lambda))\left(\muphi\Phi\left({\lambda \over
\eps}\right)\right)\, d\lambda \cr
&=\tilde f\left(\left(\muphi\Phi\right)_\eps, \mu(x)\right) \cr }
$$
However $\muphi\Phi$ now depends upon $\eps$ (although unlike 
$\Phi_\eps$ it is bounded as $\eps \to 0$) so we replace
$\OCA{0}$ by the space $\hat\OCA{0}$ of \emph{bounded paths} in
$\cD(\Real^n)$. 
\definition{Definition \defnum}
\xlabel{Ako}
The space $\hat\OCA{0}$ is the set of all smooth maps
$$\eqalign{
(0,1] &\to \cD(\Real^n) \cr
\eps &\mapsto \Phi^{(\eps)} \cr}$$
such that $\left\{\Phi^{(\eps)}: \eps \in (0,1]\right\}$ is
bounded in $\cD(\Real^n)$ and
$$
 \int \Phi^{(\eps)}(\xi) \,d\xi=1 \quad \forall \eps \in (0,1]
$$
\enddefinition

However as well as an $\eps$-dependence we see that $\muphi$ also
introduces an $x$-dependence. We therefore modify the definition of 
$\hat\OCA{0}$ as follows:
\definition{Definition \defnum}
\xlabel{Aks}
The space $\CA{0}(\Real^n)$ is defined to be the set of all $\Pex
\in \cD(\Real^n)$ where $\Pex=\muphi\Phi^{(\eps)}$ for some $\Phi^{(\eps)}
\in \hat\OCA{0}(\Real^n)$  and some  $\mu \in \Diff{\Real^n}{\Real^n}$
\enddefinition
Note that elements of $\CA{0}(\Real^n)$ satisfy the normalisation condition
$$ \int \Pex(\xi) \,d\xi=1 \quad \forall (\eps,x)\in(0,1]\times\Real^n
   \eqno\eqnum\xlabel{moment1} $$
and that this condition is preserved by $\muphi$. Unfortunately the
moment condition \eqref{mom2} is not preserved by $\muphi$. We
therefore weaken the moment condition as follows:

\definition{Definition \defnum}
The space $\CA{k}(\Real^n)$ is the space of all elements of
$\CA0(\Omega)$ which have the property 
$$ \int \xi^{\alpha} \Pex(\xi) \,d\xi=O(\eps^k)  \qquad\forall
   \alpha\in\Nat^n, \quad 1\leq|\alpha|\leq k \eqno\eqnum\xlabel{moment2} $$
\enddefinition

The diffeomorphism $\mu:\Real^n\to\Real^n$ induces a map 
$$ \eqalign{
   & \muphi: \CA0(\Real^n) \to \CA0(\Real^n) \cr
   & \muphi\Pex(\lambda) = {1\over\left|\mujac(\mmu(\mu(x)+\eps\lambda))
   \right|} \Pex \left({\mmu(\mu(x)+\eps\lambda)-x\over\eps}\right) \cr
}\eqno\eqnum\xlabel{trans.phi}
$$

\proclaim{Theorem \procnum}
$\muphi$ exhibits the following properties

\item{1.} $\Phi\in\CA{0}(\Real^n) \implies \muphi\Phi\in\CA{0}(\Real^n)$

\item{2.} $\Phi\in\CA{k}(\Real^n) \implies \muphi\Phi\in\CA{\halfk}
(\Real^n)$

\item{3.} If $\mu:\Real^n\to\Real^n$ and $\nu:\Real^n\to\Real^n$ are
diffeomorphisms then $(\nu\circ\mu)_* = \nu_*\circ\mu_*$.

\endproclaim

\demo{Proof} Refer to Colombeau and Meril (1994). \enddemo

\definition{Definition \defnum} The unbounded path $(\Pxe)$ corresponding to
$\Pex\in\CA{0}(\Real^n)$ is defined by
$$ \Pxe(\xi) = {1\over\eps^n} \Pex(\xi/\eps) $$
\enddefinition

The unbounded path may be regarded as a regularisation for the delta
distribution since $\forall\Psi\in\cD(\Real^n)$;
$$ \lim_{\eps\to0} \int \Pxe(\xi) \Psi(\xi) \,d\xi = \Psi(0) $$
hence its alternative name, the delta-net. Here the unbounded path
$(\Pxe)\notin\CA0(\Real^n)$, unlike its counterpart in Colombeau's original
theory.  A transformation law for these unbounded paths may be formed
from~\eqref{trans.phi}; Writing $\Phi'{}^{(\eps)}_{\mu(x)}=\muphi\Pex$ we
have that
$$ \eqalign{
   \Phi'_{\mu(x),\eps}(\lambda) & = {1\over\eps^n}
\Phi'{}^{(\eps)}_{\mu(x)}(\lambda) \cr
   &= {1\over\eps^n}  {1\over\left|\mujac(\mmu(\mu(x)+\lambda))
   \right|} \Pex \left({\mmu(\mu(x)+\lambda)-x\over\eps}\right) \cr
   &= {1\over\left|\mujac(\mmu(\mu(x)+\lambda))\right|} \Pxe
   \left({\mmu(\mu(x)+\lambda)-x}\right) \cr
}$$
which we shall also express as
$$ \muphi \Pxe (\lambda) = {1\over\left|\mujac(\mmu(\mu(x)+\lambda))
   \right|} \Pxe \left({\mmu(\mu(x)+\lambda)-x}\right) $$
It should be noted that in this context $\muphi$ is a map acting on the
space of  unbounded paths and not on elements of $\CA{0}$.

Generalised functions are then defined to make use of the new $\CA{k}$s;

\definition{Definition \defnum}
The space $\CE(\Real^n)$ is the set of all maps
$$ \eqalign{
   \tf: \CA0 \times \Real^n &\to \Real \cr
   (\Pex,x) & \mapsto \tf(\Pex,x) \cr
}$$
which are $C^\infty$ as functions of $x$
\enddefinition
Since the smoothing kernels now depend upon $x$, it is no longer
sufficient to require smoothness with respect to the second argument
of $\tilde f$. Instead we also require some kind of smooth dependence
upon the smoothing kernels. Colombeau and Meril require
Silva-differentiability of $\tilde f$ as a function of the
mollifier (Colombeau, 1982). An alternative, and possibly simpler 
requirement, is to demand differentiability in the sense of calculus 
on `convenient vector spaces' (Kriegl and Michor, 1997).
We next define a subalgebra of moderate functions
\definition{Definition \defnum}
$\CM(\Real^n)$ is the space of $\tf\in\CE(\Real^n)$ such that $\forall
K\compact\Real^n$, $\forall
\alpha\in\Nat^n$, $\exists N\in\Nat$ such that
$\forall\Pex\in\CA{N}(\Real^n)$,
$$ \sup_{x\in K} \left\vert D^\alpha \tf(\Pxe,x) \right\vert =
   O(\eps^{-N}) \qquad \hbox{(as $\eps\to0$)} $$
\enddefinition
\noindent
Note that this differs from the definition in Colombeau and Meril because
our $\CA{0}$ have an $x$-dependence.

We now define an ideal of negligible (null) functions
\definition{Definition \defnum}
$\CN(\Real^n)$ is the space of functions $\tf\in\CM(\Real^n)$ such that
$\forall K\compact\Real^n$, $\forall \alpha\in\Nat^n$, $\exists N\in\Nat$,
$\exists \sigma\in S$ such that $\forall\Pex\in\CA{k}(\Real^n)$ ($k\geq N)$, 
$$ \sup_{x\in K} \left\vert D^\alpha \tf(\Pxe,x) \right\vert = 
   O(\eps^{\sigma(k)}) \qquad \hbox{(as $\eps\to0$)} $$
where
$$ S = \{\, \sigma:\Nat\to\Real^+ \,|\, \sigma(k+1)>\sigma(k), \
   \sigma(k)\to\infty \,\} $$
\enddefinition

The space of generalised functions is then defined as the quotient
$$ \CG(\Real^n) =  {\CM(\Real^n) \over \CN(\Real^n) }. $$

We may then define a map
$$ \eqalign{
   \muem:\CE(\Real^n) &\to \CE(\Real^n) \cr 
   \muem\tf_\eps(\Pex,x) &= \tf_\eps(\muphi\Pex,\mu(x)) \cr
}$$
The fact that $\muphi$ maps the kernel spaces $\CA{k}(\Real^n)$ into
$\CA{\halfk}(\Real^n)$ will imply that
$$ \eqalign{
   & \muem\tf\in\CM(\Real^n) \iff \tf\in\CM(\Real^n) \cr
   & \muem\tf\in\CN(\Real^n) \iff \tf\in\CN(\Real^n)  \cr
}$$
which will induce a well defined map $[\muem]:\CG(\Real^n)\to\CG(\Real^n)$.
Furthermore if we define an embedding by
$$ \eqalign{
   \embed: \CC(\Real^n) &\to \CM(\Real^n) \cr 
   \embed(f) &= \tf \cr
   \tf(\Pex,x) &= \int f(x+\eps\xi) \Pex(\xi) \,d\xi \cr
} \eqno\eqnum\xlabel{conv}$$   
then if $f\in\CC(\Real^n)$ we have
$$ \eqalign{
   \muem\tf(\Pex,x) &= \int f(\mu(x)+\eps\lambda) \Pex\left(
   {\mmu(\mu(x)+\eps\lambda)-x\over\eps}\right) {d\lambda
   \over \left|\mujac(\mmu(\mu(x)+\eps\lambda))\right|} \cr
&=\int f(\mu(x+\eps\xi)) \Pex(\xi) \,d\xi \cr
}$$
giving us that
$$ \muem\circ\embed(f) = \embed\circ\mufun (f) $$
as we required.

We now define the algebra $\CG(\Omega)$ on open sets $\Omega \subseteq
\Real^n$. $\CM(\Omega)$ and $\CN(\Omega)$ are defined in the obvious
way by restricting $x$ to be in $\Omega$. Then
$$
\CG(\Omega)=\CM(\Omega)/\CN(\Omega)
$$

If we now only have a diffeomorphism between open sets $\mu:\Omega \to
\Omega'$ rather than the whole of $\Real^n$ we see that $\muphi:
\CA{0}(\Real^n) \to \CA{0}(\Real^n)$ is not always well defined by
\eqref{trans.phi}. However it is always well defined for $x \in
\Omega$ and sufficiently small $\eps$. Since this is all that is
needed for the theory we may define $\muphi\Pex$ for other values in
some arbitrary manner without effecting the definition of $\muem\tilde
f(\Pex, x)$. Thus a diffeomorphism $\mu:\Omega \to \Omega'$ induces a
well defined map $\muem:\CG(\Omega') \to \CG(\Omega)$. 

We now consider the embedding. Given $f \in \cD(\Omega)$ we extend
$f$ to some function $g \in \cD(\Real^n)$ such that $g|_\Omega=f$. We
then embed $g$ into $\CM(\Real^n)$ using \eqref{conv}. Again one can
show that for all $x \in \Omega$ and sufficiently small $\eps$ the
answer does not depend upon the extension. We therefore have a well
defined embedding $\embed:\cD(\Omega) \to \CG(\Omega)$. Furthermore we
still have the result that 
$$
\muem\circ \embed(f)=\embed\circ\mufun(f) \eqno\eqnum\xlabel{com}
$$
as the previous result is valid for sufficiently small $\eps$.

We are finally in a position to consider generalised functions on
manifolds. Let $M$ be a manifold with atlas ${\cal A}=\{ (\psi_\alpha,
V_\alpha): \alpha \in A\}$. Given a function $f \in \cD(M)$ we define 
$f_\alpha: \psi_\alpha(V_\alpha) \to \Real$ by
$f_\alpha=f\circ\psi^{-1}_\alpha$. The set $\{f_\alpha \}_{\alpha \in A}$
then satisfies
$$\eqalign{
f_\beta|_{\psi_\beta(V_\alpha\cap V_\beta)}
&=f_\alpha|_{\psi_\beta(V_\alpha\cap V_\beta)}
\circ\psi_\alpha\circ\psi^{-1}_\beta \cr
&=(\psi^{-1}_\beta)^*\psi_\alpha^*f_\alpha|_{\psi_\beta(V_\alpha\cap 
V_\beta)}\quad \forall \alpha,\ \beta \in A
\hbox{ with } V_\alpha \cap V_\beta \neq \emptyset \cr}
\eqno\eqnum\xlabel{trans}
$$
Conversely any set of functions $\{f_\alpha \in
\cD(\psi_\alpha(V_\alpha)) \}_{\alpha \in A}$ which satisfies
\eqref{trans} defines an element of $\cD(M)$. In exactly the same way
we make the following definition
\definition{Definition \defnum}
A generalised function $\tilde f \in \CG(M)$ is a set of generalised
functions $\{\tilde f_\alpha \in \CG(\psi_\alpha(V_\alpha))\}_{\alpha
\in A}$ which satisfies
$$
\tilde f_\beta|_{\psi_\beta(V_\alpha\cap V_\beta)}
=(\tilde\psi^{-1}_\beta)^*\tilde\psi_\alpha^*\tilde f_\alpha|_
{\psi_\beta(V_\alpha\cap 
V_\beta)}\quad \forall \alpha,\ \beta \in A
\hbox{ with } V_\alpha \cap V_\beta \neq \emptyset
$$
\enddefinition

If we define $\tilde f_\alpha = \embed_\alpha \circ f_\alpha$ where
$\embed_\alpha$ is the embedding $\embed_\alpha:\cD(\psi_\alpha(V_\alpha))
\to \CG(\psi_\alpha(V_\alpha))$, then one has an embedding
$$
\embed:\cD(M) \to \CG(M)
$$

\head
\secnum.
Smoothing of vector and covector fields
\endhead
We now have a  coordinate invariant theory of generalised functions
at the level of scalars. We would like to extend this theory to enable
vectors, covectors and, ultimately, multi-index tensors to be defined as
generalised functions whose transformation laws will coincide with those of
distributions.

We shall first consider the smoothing of covector field $\omega$, which may
be represented in both coordinate systems $x\in\Omega$ and $x'\in\Omega'$ as
$$ \omega=\omega_a(x) dx^a = \omega'_a(x') dx'{}^a, $$
with the  components $\omega_a$ and $\omega'_a$  related by
$\omega_a=\mufun\omega'_a$ where 
$$ \eqalign{
   \mufun: \CC\st01(\Omega') &\to \CC\st01(\Omega), \cr
   \mufun \omega'_a (x) &= \mux{b}{a}(x) \omega'_b(\mu(x)). \cr} $$

We could use a componentwise smoothing,
$$ \eqalign{
   \embed: \CC^0_1(\Omega) &\to \CE^0_1(\Omega) \cr
   \embed(\omega) &= \tomega \cr
   \tomega_a(\Phi,x) &=\int \omega_a(x+\eps\xi) \Pex(\xi) \, d\xi \cr}
   \eqno\eqnum\xlabel{componentwise} $$
and relate the smoothings by $\muem:\CE^0_1(\Omega')\to\CE^0_1(\Omega)$
$$ \muem \tomega'_a (\Phi,x) = \mux{b}{a}(x) \tomega'_b (\muphi\Phi,x) $$
in which case we would have that
$$ \eqalign{
  \tomega_a(\Phi,x)   
   &= \int \mux{b}{a}(x+\eps\xi) \omega'_b \bigl(\mu(x+\eps\xi)\bigr)
   \Pex(\xi) \,d\xi \cr
   &= \int \mux{b}{a}\Bigl(\mmu\bigl(\mu(x)+\eps\lambda\bigr)\Bigr)
   \omega'_b\bigl(\mu(x)+\eps\lambda\bigr) \muphi\Pex(\lambda) \,d\lambda
   \cr
}$$
and
$$ \muem \tomega'_a(\Phi,x) = \int \mux{b}{a}(x) \omega'_b(\mu(x)+\eps\lambda)
   \muphi\Pex(\lambda)  \, d\lambda $$
If it is the case that $\omega'_a$ is $C^\infty$, we can expand $\mux{b}{a}$ 
and $\omega'_a$ in powers of $\eps\lambda$  which will reveal that
$$ \tomega_a(\Phi,x)-\muem\tomega'_a(\Phi,x)=O(\eps^k) \qquad \hbox{for }
   \muphi\Phi\in\CN(\Omega')$$
so $\tomega_a(\Phi,x)$ and $\muem\tomega'_a$ may be regarded as
representing the same element of $\CG(\Omega)$. On the other hand if
$\omega'_a$ admitted only a finite level of differentiability, then
$\tomega_a$ and $\muem \tomega'_a$ would not be equivalent as elements of
$\CG(\Omega)$ although they may well be equivalent at the level of association,
given that the level of differentiability was high enough.

A possible remedy for this problem is to introduce smoothing kernels with
indices, thus we replace~\eqref{componentwise} by
$$ \tomega^a(x) = \int \omega_b(x+\eps\xi) \Pexcov{a}{b}(\xi) \,d\xi $$
where
$$ \eqalign{
   & \int \Pexcov{a}{b}(\xi) \,d\xi = \kron{b}{a}+O(\eps^k) \cr
   & \int \xi^\alpha \Pexcov{a}{b}(\xi) \,d\xi = O(\eps^k) \qquad
     \alpha\in\Nat^n, \quad 1\leq|\alpha|\leq k \cr
} \eqno\eqnum\xlabel{moment.cov}$$
and define
$$ \muphi \Pexcov{a}{b}(\lambda) = {\mmux{c}{a}(x)
   \mux{b}{d}(\mmu(\mu(x)+\eps\lambda)) \over \left|
   \mujac(\mmu(\mu(x)+\eps\lambda)) \right | }\Pexcov{c}{d}
   \left({\mmu(\mu(x)+\eps\lambda)-x \over \eps} \right)
   \eqno\eqnum\xlabel{trans.phicov} $$
where
$$ \mmux{a}{b}(x)=\mmu^a_b(\mu(x)) $$

\proclaim{Proposition \procnum}
The moment conditions~\eqref{moment.cov} are preserved under $\muphi$ in
that 
$$ \eqalign{
   & \int \muphi\Pexcov{a}{b}(\lambda) \,d\lambda = \kron{b}{a}+O(\eps^k)
   \cr 
   & \int \lambda^\alpha \muphi\Pexcov{a}{b}(\lambda) \,d\lambda =
   O(\eps^\halfk) \qquad \alpha\in\Nat^n, \quad 1\leq|\alpha|\leq \halfk 
   \cr  
}$$
\endproclaim

\demo{Proof}
The first condition holds since
$$ \eqalign{
   \int \Pexcov{a}{b}(\lambda)\, d\lambda
   &= \int \mmux{c}{a}(x) \mux{b}{d}(x+\eps\xi) \Pexcov{c}{d}(\xi) \,d\xi
   \cr  
   &= \mmux{c}{a}(x) \mux{b}{d}(x) \int \Pexcov{a}{b}(\xi)\,d\xi +
   \sum_{|\alpha|=1}^k D^\alpha\mmux{c}{a}(x) \mux{b}{d}(x)
   {\eps^{|\alpha|}\over 
   |\alpha|!}  \int \xi^\alpha \Pexcov{c}{d}(\xi) \,d\xi \cr
   &\phantom{=} \quad +\sum_{|\alpha|=k+1}
   D^\alpha\mmux{c}{a}(x+\eps\theta\xi) \mux{b}{d}(x) {\eps^{|\alpha|}\over
   |\alpha|!} \int \xi^\alpha \Pexcov{c}{d}(\xi) \,d\xi \cr
   &= \kron{b}{a} + O(\eps^{k+1})
}$$
A similar calculation (See Colombeau and Meril, 1994) will show that the
second condition is preserved with $k$ replaced by $\halfk$.
\enddemo

\proclaim{Proposition \procnum}
$\muem \tomega'_a = \tomega_a $
\endproclaim

\demo{Proof}
$$ \eqalign{
  \tomega_a(\Phi,x) &= \int \mux{c}{b}(x+\eps\xi) \omega'_c(\mu(x+\eps\xi))
  \Pexcov{a}{b}\left(\xi\right) \,d\xi \cr
  &= \int \mux{c}{b}(\mmu(\mu(x)+\eps\lambda)) \omega'_c(\mu(x)+\eps\lambda) 
  \Pexcov{a}{b}\left({\mmu(\mu(x)+\eps\lambda)-x \over \eps}
  \right) \,d\lambda \cr
  &= \int \mux{b}{a}(x) \omega'_c(\mu(x)+\eps\lambda)
  \muphi\Pexcov{b}{c}(\eps,\mu(x),\lambda) \, d\lambda \cr
  &= \muem\tomega'_a(\Phi,x)
}$$
\enddemo

The need for smoothing kernels with indices is clear, because we are
integrating the coefficients of the covector field at the point $x+\eps\xi$
to give a covector at the point $x$. The kernel $\Phicov{a}{b}$ therefore
has the effect of transporting a covector at the point $x+\eps\xi$ to the
point $x$. Rather than considering general kernels with indices
$\Phicov{a}{b}$ we shall consider those which have the form
$$ \Pexcov{a}{b}(\xi)=\trcov{a}{b}(x,x+\eps\xi)\Pex(\xi) $$
where 
$$ \eqalign{ 
   \trcov{}{} : \Omega\times\Omega &\to \cotang(\Omega)\times \tang(\Omega) \cr
   (x,y) &\mapsto \trcov{a}{b}(x,y) \cr
}$$
is a smooth map with $\trcov{a}{b}(x,x)=\kron{b}{a}$ and
$\Pex\in\CA0(\Real^n)$. The respective transformation laws~\eqref{trans.phi}
and~\eqref{trans.phicov} for $\Pex$ and $\Phicov{a}{b}$ will imply that
$\trcov{a}{b}(x,y)$ has to transform as
$$ \muphi \trcov{a}{b}(\mu(x),\mu(y)) = \eta_a^c(x) \mux{b}{d}(y)
   \trcov{c}{d} (x,y) $$
so $\trcov{a}{b}$ may be regarded as components of a covector (with index
$a$) located at $x$ and a vector (with index $b$) located at $y$.

A similar procedure could be applied to smooth a vector
$$ X=X^a(x)\pbyp{}{x^a}=X'{}^b(x)\pbyp{}{x'{}^b} $$
with the  components $X^a$ and $X'{}^a$  related by
$X^a=\mufun X'{}^a$ where 
$$ \eqalign{
   \mufun: \CC\st10(\Omega') &\to \CC\st10(\Omega), \cr
   \mufun X'{}^a (x) &= \mmux{a}{b}(x) X'{}^b(\mu(x)). \cr} $$
In this case we use a smoothing of the form
$$ \tX^a(x) = \int X^b(x+\eps\xi) \Pexvec{a}{b}(\xi) \, d\xi $$
By allowing $\Phivec{a}{b}$ to transform as
$$ \muphi \Pexvec{a}{b}(\lambda) = {\mux{a}{c}(x)
   \mmux{d}{b}(\mmu(\mu(x)+\eps\lambda)) \over \left|
   \mujac(\mmu(\mu(x)+\eps\lambda)) \right |} \Pexvec{c}{d}
   \left({\mmu(\mu(x)+\eps\lambda)-x \over \eps} \right) $$
we are able to define a map $\muem$ by
$$ \muem\tomega'{}^a(\Phi,x) = \mmux{a}{b}(x) \tomega'{}^b(\muphi\Phi,\mu(x)) $$
which will satisfy
$$ \muem\tX'{}^a = \tX^a $$
This time $\Phivec{a}{b}$ transports a vector at $x+\eps\xi$ to a vector at
$x$, we may therefore consider kernels $\Phivec{a}{b}$ which admit the form
$$ \Pexvec{a}{b}(\xi)=\trvec{a}{b}(x,x+\eps\xi)\Pex(\xi) $$
where $\Pex\in\CA0(\Real^n)$ and the map
$$ \eqalign{ \transp : \Omega\times\Omega &\to \tang(\Omega) \times \cotang(\Omega) \cr
   (x,y) & \mapsto \trvec{a}{b}(x,y) \cr} $$
is smooth with $\trvec{a}{b}(x,x)=\kron{a}{b}$ and transforms as
$$ \muphi \trvec{a}{b}(\mu(x),\mu(y)) = \mux{a}{c}(x) \mmux{d}{b}(y) \trvec{c}{d}
   (x,y) \eqno\eqnum\xlabel{trans.gamma} $$

We shall formalise our transport operators $\trvec{a}{b}$ and
$\trcov{a}{b}$ by defining the following space of transport operators;

\definition{Definition \defnum}
The space $\CT(\Omega)$ is defined to be the set of maps
$$ \eqalign{
   \transp :\Omega\times\Omega &\to \tang(\Omega) \times \cotang(\Omega) \cr
   (x,y) &\mapsto \trvec{a}{b}(x,y) \cr
}\eqno\eqnum\xlabel{connector.space}$$   
which are smooth and are  such that $\trvec{a}{b}(x,x)=\kron{a}{b}$.
\enddefinition

In this way we may simultaneously write down smoothings of vectors and
covectors as functions $\CA0(\Real)\times\CT(\Omega)\times\Omega \to \Real$
$$ \eqalign{
   \tX^a(\Phi,\transp,x) &= \int X^b(x+\eps\xi) \trvec{a}{b}(x,x+\eps\xi)
   \Phi(\xi) \, d\xi \cr
   \tomega_a(\Phi,\transp,x) &= \int \omega_b(x+\eps\xi)
   \trcov{a}{b}(x,x+\eps\xi) \Phi(\xi) \,d\xi \cr
} \eqno\eqnum\xlabel{embed}$$
where
$$ \trvec{a}{c} \trcov{b}{c} = \kron{c}{b}. \eqno\eqnum\xlabel{connector.inv}$$
An important consequence of this arrangement is that $\transp$ will preserve
contractions of covectors with vectors; suppose that
$$ \eqalign{ X^a(x) &= \trvec{a}{b}(x,y) X'{}^b(y) \cr \omega_a(x) &=
   \trcov{a}{b} (x,y) \omega'_b(y) \cr }$$
then
$$ \eqalign{ \omega_a(x) X^a(x) &= \trcov{a}{b}(x,y)\trvec{a}{c}(x,y)
   \omega'_a(y) X'{}^a(y) \cr &= \omega'_a(y) X'{}^a(y) \cr} $$

\head
\secnum.
Remarks on the transport operator
\endhead
We now show that the derivative of the transport operator defines a
connection and that conversely a connection defines a transport operator
(in a normal neighbourhood).  We first consider rules for differentiating
$\transp$.  We may write
$$ \pbyp{\trcov{a}{b}}{x^c}(x,y) = \con{d}{ca}(x) \trcov{b}{d}(x,y) $$
using the fact that
$$ \eqalign{
   & \trvec{a}{b}(x,y) \trvec{b}{c}(y,x)=\kron{a}{c} \cr
   & \trvec{a}{c}(x,y) \trcov{b}{c}(x,y)=\kron{a}{b} \cr
}$$
we also have
$$ \eqalign{
   \pbyp{\trcov{a}{b}}{y^c}(x,y) &= -\con{b}{cd}(y) \trcov{a}{d}(x,y) \cr
   \pbyp{\trvec{a}{b}}{x^c}(x,y) &= -\con{a}{cd}(x) \trvec{d}{b}(x,y) \cr 
   \pbyp{\trvec{a}{b}}{y^c}(x,y) &= \con{d}{cb}(y) \trvec{d}{a}(x,y) \cr  
}$$

\proclaim{Proposition \procnum}
$\con{a}{bc}$ transforms as a connection
\endproclaim

\demo{Proof}
The connector $\trcov{a}{b}$ transforms as
$$ \trcovp{a}{b}(x',y') = \pbyp{x^c}{x'{}^a} \pbyp{y'{}^b}{y^d}
   \trcov{c}{d}(x,y) $$
so
$$ \pbyp{\trcov{a}{b}}{x'{}^c} = \pbypmix{x^e}{x'{}^a}{x'{}^c}
   \pbyp{y'{}^b}{y^d} \trcov{e}{d} + \pbyp{x^e}{x'{}^a} \pbyp{x^f}{x'{}^c}
   \pbyp{y'{}^b}{y^d} \pbyp{\trcov{e}{d}}{x^f} $$
this implies that
$$ \conp{d}{ca} \trcovp{d}{b} = \pbypmix{x^e}{x'{}^a}{x'{}^c}
   \pbyp{y'{}^b}{y^d} \trcov{e}{d} + \pbyp{x^f}{x'{}^c} \pbyp{y'{}^b}{y^d}
   \con{g}{fe}\trcov{g}{d} $$
and hence
$$ \conp{e}{ca} \pbyp{x^g}{x'{}^e}\pbyp{y'{}^b}{y^d}\trcov{g}{d} =
   \pbypmix{x^g}{x'{}^a}{x'{}^c} \pbyp{y'{}^b}{y^d} \trcov{g}{d} +
   \pbyp{x^f}{x'{}^c} \pbyp{y'{}^b}{y^d} \con{g}{fe}\trcov{g}{d} $$
on multiplying this expression throughout by
$$ \pbyp{y^p}{y'{}^b} \pbyp{x'{}^r}{x^q} \trvec{q}{p} $$
one obtains the standard connection transformation law
$$ \conp{a}{bc} = \pbyp{x'{}^a}{x^d}\pbyp{x^e}{x'{}^b}\pbyp{x^f}{x'{}^c}
   \con{d}{ef} + \pbyp{x'{}^a}{x'{}^d} \pbypmix{x^d}{x'{}^b}{x'{}^c} $$
\enddemo

If $\gamma$ is a connection then one can always choose a neighbourhood
$U_x$ of any point $x\in\Omega$ such that the normal coordinates at $x$ are
well defined. Moreover one can choose $U_x$ to be simply convex; that is
that $U_x$ is also a normal neighbourhood for any point $y\in U_x$.  In a
simply convex neighbourhood one can define $\trvec{a}{b}(x,y)$ by parallel
propagation with respect to $\gamma$ along the unique geodesic connecting
$x$ and $y$, so in geodesic coordinates $\trvec{a}{b}(x,y)=\kron{a}{b}$

In general if we allow any connection $\gamma$, the sets $\Omega$ in the
atlas will not be normal neighbourhoods. However since $\Pex$ has compact
support, the integration will be confined to a normal neighbourhood for
sufficiently small $\eps$. Since we are only interested in results for
sufficiently small $\eps$ we could also work with $\gamma$ in place of
$\transp$. However once we consider tensor fields on manifolds, rather than
open sets $\Omega$, it is much less restrictive to work with a connection
$\gamma$ rather than a global connector $\Gamma$. Even when working with
subsets $\Omega$ of $\Real^n$ it is preferable to work with connections on
$\Real^n$ rather than connectors on $\Omega$, since the set of algebras
then has the structure of a (fine) sheaf. If one does this, then one
defines the embedding by \eqref{embed}, where $\Gamma$ is defined by
parallel transport along geodesics in a normal neighbourhood and
arbitrarily elsewhere. Since one is in a normal neighbourhood for
sufficiently small epsilon, the embedding into the algebra will be well
defined.

\head
\secnum.
General tensors
\endhead
Having defined procedures for smoothing covectors and vectors in such a way
that $\muem\circ\embed' = \embed\circ \mufun$, it is now clear how to go
about extending this smoothing operation to more general tensors; Suppose
$T$ is a tensor field of type $(p,q)$ whose components in the two
coordinate systems $x\in\Omega$ and $x'\in\Omega'$ are related by
$T'\gt{a}{b}{c}{d} = \mufun T\gt{a}{b}{c}{d}$ where
$$ \mufun T'\gt{a}{b}{c}{d}(x) = \mmux{a}{e}(x) \ldots \mmux{b}{f}(x)
   \mux{g}{c}(x) \ldots \mux{h}{d}(x) T'\gt{e}{f}{g}{h}(\mu(x)) $$
then we may define a smooth tensor field by
$$ \eqalign{
   &\tT\gt{a}{b}{c}{d}(\Phi,\transp,x)=\int
   T\gt{e}{f}{g}{h}(x+\eps\xi) \Pex\ud{a\ldots b}{c\ldots d}
   \du{e\ldots g}{g\ldots h}(\xi) \, d\xi \cr
   &\Pex\ud{a\ldots b}{c\ldots d}\du{e\ldots g}{g\ldots h}(\xi) =
   \trvec{a}{e}(x+\eps\xi) \ldots \trvec{b}{f}(x+\eps\xi)
   \trcov{c}{g}(x+\eps\xi) \ldots \trcov{d}{h}(x+\eps\xi) \Pex(\xi) \cr
}\eqno\eqnum\xlabel{smooth.tensor}$$
and a map $\muem$ by
$$ \muem\tT'\gt{a}{b}{c}{d}(\Phi,\transp,x) = \mmux{a}{e}(x) \ldots
   \mmux{b}{f}(x) \mux{g}{c}(x) \ldots \mux{h}{d}(x)
   \tT'\gt{e}{f}{g}{h}(\muphi\Phi,\muphi\transp,\mu(x)) 
\eqno\eqnum\xlabel{tensortrans}$$
which will satisfy
$$ \tT\gt{a}{b}{c}{d} = \muem \tT'\gt{a}{b}{c}{d} \eqno\eqnum\xlabel{tensortrans2}
$$

We are now in a position to formally define our generalised tensor fields
on $\Omega$. The kernel space $\CA{0}(\Real^n)$ is defined as in
Definition~\xref{Aks} and $\CK(\Real^n)$ is the space of smooth
connections on $\Real^n$. We first define the base algebra
$\CE\st{p}{q}(\Omega)$ 

\definition{Definition \defnum}
Let $\CE\st{p}{q}(\Omega)$ be the set of all maps
$$ \eqalign{
   \tT: \CA0 \times \CK \times \Omega &\to \Real^{p+q} \cr
   (\Phi,\gamma, x) & \mapsto \tT\gt{a}{b}{c}{d}(\Phi,\gamma,x) \cr
}$$
such that $x \mapsto \tT\gt{a}{b}{c}{d}(\Phi,\gamma,x)$ is $C^\infty$.
\enddefinition
As in the case of scalar fields we require some smooth dependence upon the
smoothing kernel and connection to ensure smooth dependence in $x$.  As
usual we restrict to those of moderate growth
\definition{Definition \defnum}
$\CE\st{p}{q,M}(\Omega)$ is the space of tensors
$\tT\in\CE\st{p}{q}(\Omega)$ such that $\forall K\compact\Omega$, $\forall
\alpha\in\Nat^n$, $\exists N\in\Nat$ such that
$\forall\Pex\in\CA{N}(\Real^n)$, $\forall\gamma\in\CK$,
$$ \sup_{x\in K} \left\Vert D^\alpha \tT(\Pxe,\gamma,x) \right\Vert =
   O(\eps^{-N}) \qquad \hbox{(as $\eps\to0$)} $$
\enddefinition

We define an ideal of negligible (null) functions

\definition{Definition \defnum}
$\CN\st{p}{q}(\Omega)$ is the space of tensors
$\tT\in\CE\st{p}{q,M}(\Omega)$ such that $\forall K\compact\Omega$,
$\forall
\alpha\in\Nat^n$, $\exists N\in\Nat$, $\exists \sigma\in S$ such that
$\forall\Pex\in\CA{k}(\Real^n)$ ($k\geq N)$, $\forall\gamma\in\CK$,
$$ \sup_{x\in K} \left\Vert D^\alpha \tT(\Pxe,\gamma,x) \right\Vert = 
   O(\eps^{\sigma(k)}) \qquad \hbox{(as $\eps\to0$)} $$
where
$$ S = \{\, \sigma:\Nat\to\Real^+ \,|\, \sigma(k+1)>\sigma(k), \
   \sigma(k)\to\infty \,\} $$
\enddefinition
We then define our generalised function space as a quotient
\definition{Definition \defnum}
$$ \CG\st{p}{q}(\Omega) =  {\CE\st{p}{q,M}(\Omega) \over \CN\st{p}{q}(\Omega) } $$
\enddefinition

One may now define generalised tensor fields on manifolds as a collection
of generalised tensor fields on $\psi_\alpha(V_\alpha)$ which transform in
the appropriate way under $(\tilde\psi_\beta^{-1})^*(\tilde\psi_\alpha)^*$
using~\eqref{tensortrans}. Because of \eqref{tensortrans2} we also have a
well defined embedding of $\cD^p_q(M)$ into $\CG\st{p}{q}(M)$.

Throughout the paper we have been using distributions that may be defined
as locally integrable functions. It is possible to extend our results to
cover the smoothing of more general tensors.  We first recall how
distribution theory is formulated in a coordinate invariant manner; Suppose
$\mu:\Omega\to\Omega'$ is a $C^\infty$ diffeomorphism, then we may define
$\tCD^q_p(\Omega)$ to be the space of $C^\infty$ multi-index functions with
compact support which transform as type $(q,p)$ densities with weight $+1$
under
$$ \muphi\Psi\gt{c}{d}{a}{b}(\mu(x)) = {1\over \left| J_\mu(x)\right|}
   \mmux{c}{g}(x) \ldots \mmux{d}{h}(x) \mux{e}{a}(x) \mux{f}{b}(x)
   \Psi\gt{c}{d}{a}{b}(x) $$ 
We simply define type $(p,q)$ tensor distributions as elements of the dual
space $\tCD'{}^{p}_{q}(\Omega)$ which then admit a transformation law
$\mufun:\tCD'{}^{p}_{q}(\Omega')\to\tCD'{}^{p}_{q}(\Omega)$ of the form
$$ \tprod{\mufun T'}{\Psi} = \tprod{T}{\mu_*\Psi} $$
This immediately becomes evident in case of $T$ being a locally integrable
tensor field, for the integral
$$ \tprod{T}{\Psi} = \int T\gt{a}{b}{c}{d}(x)
   \Psi\gt{c}{d}{a}{b}(x) \,dx $$
is coordinate invariant.

We now define
$$ \eqalign{
   \embed:\cD'(\Omega) &\to\CM(\Omega) \cr
   \embed(T) &= \tT \cr
}$$
where
$$ \eqalign{
   &\tT\gt{a}{b}{c}{d}(\Phi,\gamma,x) =
   \tprod{T}{\Xi_x\gt{a}{b}{c}{d} \tau_x \Pxe} \cr
   &( \Xi_x\gt{a}{b}{c}{d})\gt{g}{h}{e}{f}(y) = \trvec{a}{e}(x,y)
   \ldots \trvec{b}{f}(x,y) \trcov{c}{g}(x,y) \ldots \trvec{d}{h}(x,y) \cr
   &\tau_x(y)= x-y \cr
}$$

It should be noted that the test function $\Xi_x\gt{a}{b}{c}{d} \tau_x
\Pxe$ behaves as a type $(p,q)$ tensor with respect to $x$ and that the
translated object $\tau_x\Pxe$ will transform as
$$ \tau_{\mu(x)}\muphi\Pxe(\lambda)={1\over \left| \mujac(\mmu(\lambda))
   \right|} \tau_x\Pxe (\mmu(\lambda)) $$

\proclaim{Proposition \procnum}
If $T\in\cD'\st{p}{q}(\Omega')$ then $\muem\tT'=\widetilde{\mufun T'}$.
\endproclaim

\demo{Proof}
$$ \eqalign{
   \muem \tT'\gt{a}{b}{c}{d}(\Phi,\gamma,x)
   &= \tT'\gt{a}{b}{c}{d}(\muphi\Phi,\mu_*\gamma,\mu(x)) \cr
   &= \tprod{T'}{\muphi(\Xi_x \tau_x\Pxe)\gt{a}{b}{c}{d}} \cr
   &= \mmux{a}{e}(x) \ldots \mmux{b}{f}(x) \mux{g}{c}(x) \ldots
      \mux{h}{d}(x) \tprod {\mufun T'}{ \Xi_x\gt{e}{f}{g}{h}
      \tau_x\Pxe} \cr
   &= \mmux{a}{e}(x) \ldots \mmux{b}{f}(x) \mux{g}{c}(x) \ldots
      \mux{h}{d}(x) \widetilde{\mufun T'}
      \gt{e}{f}{g}{h}(\Phi,\gamma,x) \cr
}$$
\enddemo
This result enables us to define an embedding of $\cD'\st{p}{q}(M)$
into $\CG\st{p}{q}(M)$.

Since elements of the generalised function space $\CG\st{p}{q}(\Omega)$ are
represented by smooth functions, we may define tensor operations on these
generalised functions in the usual way. In particular we are able to define
the following;

\item{1.} Tensor products;
Suppose $[\tS']\in\CG{}\st{p}{q}(\Omega')$ and
$[\tT']\in\CG{}\st{r}{s}(\Omega')$ are type $(p,q)$ and type $(r,s)$
tensors respectively then we may define $[\tS'\otimes \tT'] \in
\CG\st{p+r}{q+s}(\Omega')$ by
$$ (\tS'\otimes \tT')\gt{a}{b}{c}{d} \gt{e}{f}{g}{h}(\Phi',\gamma',x') =
   \tS'\gt{a}{b}{c}{d}(\Phi',\gamma',x)
   \tT'\gt{e}{f}{g}{h}(\Phi',\gamma',x') $$
the resulting object $[\tS'\otimes\tT']$ will transform as a type
$(p+r,q+s)$ tensor in that
$$ \muem (\tS'\otimes\tT') = \muem \tS' \otimes \muem \tT' $$

\item{2.} Contractions;
Suppose $\tT'\in\CG\st{p+1}{q+1}(\Omega')$ is a type $(p,q)$ tensor then we
may define $\tS'\in\CG{}\st{p}{q}(\Omega')$ by
$$ \tS'\gt{a}{b}{c}{d}(\Phi',\gamma',x') =
   \tT'\gt{a}{be}{c}{de}(\Phi',\gamma',x') $$ 
$[\tS']$ will transform as a type $(p,q)$ tensor because
$$ \muem\tS'\gt{a}{b}{c}{d} = \muem\tT'\gt{a}{be}{c}{be} $$

\item{3.} Differentiation;
Suppose that $[\tT']\in\CG{}\st{p}{q}(\Omega')$ then we may define
$[\partial\tT']\in\CG^p_{q+1}(\Omega')$ by
$$ \partial'\tT'\gt{a}{b}{c}{de}(\Phi',\gamma',x')=
   \pbyp{\tT'\gt{a}{b}{c}{d}}{x'{}^e}(\Phi',\gamma',x') $$
In general $[\partial'\tT']$ will not transform as a tensor for we do not
necessarily have
$$ \muem(\partial'\tT') = \partial'(\muem\tT') $$
However, if we are able to define a connection
$[\tGamma'{}^a_{bc}]\in\CG^1_2(\Omega')$ that transforms as
$$ \muem\tGamma'\st{a}{bc}(\Phi,\gamma,x) = \mmux{a}{d}(x) \mux{e}{b}(x)
   \mux{f}{c}(x) \tGamma'\st{d}{ef}(\muphi\Phi,\mu_*\gamma,\mu(x)) +
   \mmux{a}{d}(x) \mux{e}{b,c}(x) $$
then we are able to define a covariant derivative
$[\nabla'\tT']\in\CG^p_{q+1}(\Omega')$ by
$$ \nabla\tT'\gt{a}{b}{c}{de} = \pbyp{}{x'{}^e}\tT'\gt{a}{b}{c}{d} +
   \tGamma'\st{a}{ef} \tT'\gt{f}{b}{c}{d}+ \cdots + \tGamma'\st{b}{ef}
   \tT'\gt{a}{f}{c}{d} - \tGamma'\st{f}{ec} \tT'\gt{a}{b}{f}{d}- \cdots
   -\tGamma'\st{f}{ed} \tT'\gt{a}{b}{c}{f} $$
which will satisfy
$$ \muem(\nabla'\tT') = \nabla'(\muem\tT') $$

\item{4.} Lie derivatives.
For $[\tX']\in\CG^0_1(\Omega')$ and $[\tT']\in\CG\st{p}{q}(\Omega)$, the
Lie derivative $[\Lie'_{\tX'}\tT]\in\CG\st{p}{q}{\Omega}$ may be defined as
$$ \Lie'_{\tX'}\tT'\gt{a}{b}{c}{d} = \tX'{}^e \tT'\gt{a}{b}{c}{de} -
  \tX'\st{a}{,f} \tT'\gt{f}{b}{c}{d} - \cdots -
  \tX'\st{b}{,f}\tT\gt{a}{f}{c}{d} + \tX'\st{f}{,c} \tT'\gt{a}{b}{f}{d}
  +\cdots+ \tX'\st{f}{,d}\tT'\gt{a}{b}{c}{f} $$
which will satisfy
$$ \muem \Lie'_{\tX'}\tT' = \Lie_{\muem X'}(\muem\tT') $$

The linear tensor operations such as addition, symmetrisation and
antisymmetrisation of indices also extend to our generalised function
valued tensors in a natural way.

\head
\secnum.
Differentiation
\xlabel{derivs}
\endhead
A $C^\infty$ tensor field $T$ may be embedded in to
$\CG\st{p}{q}(\Omega)$ in two ways;

\item{1.} By smoothing; $T \mapsto \tT$
\item{2.} By defining an element of $\CE\st{p}{q,M}(\Omega)$ which is
independent of $\Phi$ and $\gamma$; $\tS(\Phi,\gamma,x)=T(x)$

\noindent By defining $\CG\st{p}{q}(\Omega)$ as a quotient group we are
able to guarantee that $\tS$ and $\tT$ represent the same generalised
function.  Here we shall verify that this is also the case for derivatives
in that if $T\in C\st{p,\infty}q(\Omega)$ then it is the case that
$\tT\gt{a}{b}{c}{d,e}$ and $T\gt{a}{b}{c}{d,e}$ represent the same
generalised function.

We begin with scalars; suppose $f\in C^\infty(\Omega)$ then we have, 
$$ \eqalign{
   \tf_{,a}(x) &= \int f_{,a}(x+\eps\xi) \Pex(\xi) \,d\xi + \int
   f(x+\eps\xi) \Pex{}_{,a}(\xi) \, d\xi \cr	
   \widetilde{f_{,a}}(x) &= \int f_{,a}(x+\eps\xi) \Pex(\xi) \,d\xi\cr 
}$$
where $\Pex{}_{,a}$ is the partial derivative of the operator $\Pex$ with
respect to $x^a$.

We now expand the difference $\tf_{,a}-\widetilde{f_{,a}}$ as a Taylor series
(for some $\theta\in[0,1]$);
$$ \eqalign{
   \tf_{,a}-\widetilde{f_{,a}} &= \sum_{|\alpha|=0}^k {\eps^{|\alpha|}\over
   |\alpha|!} D^\alpha f (x) \int \xi^\alpha \Pex{}_{,a}(\xi)\,d\xi \cr
   & \quad + \!\!\! \sum_{|\alpha|=k+1} \!\! {\eps^{k+1}\over
   (k+1)!} D^\alpha f (x+\eps\theta\xi) \int \xi^\alpha
   \Pex{}_{,a}(\xi)\,d\xi \cr
}$$
The first term on the right-hand side may be simplified by differentiating
the moment conditions~\eqref{moment1} and ~\eqref{moment2}.
$$ \eqalign{
   & \int \Pex{}_{,a}(\xi) \,d\xi = 0 \cr
   & \int \xi^\alpha \Pex{}_{,a}(\xi) \,d\xi =O(\eps^k), \qquad
   1\leq|\alpha|\leq  
   q \cr 
}\eqno\eqnum\xlabel{diff.moments}$$
which shows that it is $O(\eps^k)$, where as the second term is clearly
$O(\eps^{k+1})$. This shows that
$\tf_{,a}-\widetilde{f_{,a}}\in\CN(\Omega)$, so that $\tf_{,a}$ and
$\widetilde{f_{,a}}$ represent the same generalised function.

If $f$ had only a finite level of differentiability then
$\widetilde{f_{,a}}$ and $\tf_{,a}$ would represent different generalised
functions.  (The difference would be $O(\eps^{m+1})$ for $f\in C^m$, which
would then make $\widetilde{f'_{,a}}$ and $\tf_{,a}$ equivalent at the
level of association).  This highlights an important difference from
Colombeau's original theory in which differentiation of distributions
manifestly commutes with smoothing. This is a price we have to pay for
introducing smoothing kernels that are dependent on $x$.

The next step is to extend this to multi-index tensors; an added
complication being that the transport operator $\transp$ will depend on
$x$, so differentiating it will introduce extra terms.

We shall use the notation $T{}_{|e}$ to denote covariant differentiation
with respect to the connection $\con{a}{bc}$ Without loss of generality we
shall consider the differentiation of vectors (the differentiation of
covectors and more general tensors is achieved by adding the relevant
terms), so for vectors we have
$$ X^a{}_{|b} = X^a_{,b} + \con{a}{bc} X^c $$

\proclaim{Lemma \procnum}
$$[\tX^a{}_{|b}(\Phi,\gamma,x)] = \left[ \int X^c{}_{|b} 
(x+\eps\xi)\trvec{a}{c}(x,x+\eps\xi) \Pex(\xi) \,d\xi  \right]$$
\endproclaim

\demo{Proof}
$$ \eqalign{
   \tX^a{}_{,b}(\Phi,\gamma,x) &= \int X^c{}_{,b}(x+\eps\xi)
   \trvec{a}{c}(x,x+\eps\xi) \Pex(\xi) \, d\xi \cr
   &\phantom{=} \quad + \int X^c(x+\eps\xi) \left( -\con{a}{bd}(x)
   \trvec{d}{c}(x,x+\eps\xi) \right.\cr
   &\phantom{=} \quad \qquad \left.+ \con{d}{bc}(x+\eps\xi)
   \trvec{a}{d}(x,x+\eps\xi) \right) \Pex(\xi) \,d\xi \cr
   &\phantom{=} \quad + \int X^c(x+\eps\xi) \trvec{a}{c}(x,x+\eps\xi)
   \Pex{}_{,b}(\xi)\, d\xi \cr
   &= \int X^c{}_{|b} (x+\eps\xi) \trvec{a}{c}(x,x+\eps\xi) \Pex(\xi)
   \,d\xi \cr
   &\phantom{=} \quad - \con{a}{bd}(x) \int X^c(x+\eps\xi)
   \trvec{d}{c}(x,x+\eps\xi) \Pex(\xi) \, d\xi \cr
   &\phantom{=} \quad + \int X^c(x+\eps\xi) \trvec{a}{c}(x,x+\eps\xi)
   \Pex{}_{,b}(\xi)\, d\xi \cr
}$$
By Taylor expanding and using the moment conditions~\eqref{diff.moments} it
may be shown that for $\Phi\in\CA{k}(\Omega)$,
$$ \int X^c(x+\eps\xi) \trvec{a}{c}(x,x+\eps\xi)
   \Pex{}_{,b}(\xi)\, d\xi = O(\eps^k) $$
therefore
$$ [\tX^a{}_{|b}(\Phi,\gamma,x)] = \left[ \int X^c{}_{|b} 
(x+\eps\xi)\trvec{a}{c}(x,x+\eps\xi) \Pex(\xi) \,d\xi  \right]$$
\enddemo

\proclaim{Proposition \procnum}
Let $X^a$ and $Y^a$ be smooth vector fields then covariant 
differentiation of $X^a$ with respect to $\con{a}{bc}$ in the $Y^a$ direction
commutes with the embedding in the sense that;
$$\widetilde{Y^bX^a{}_{|b}}=\tilde Y^b\tX^a{}_{|b}.$$
\endproclaim

\demo{Proof}
This follows from the above lemma together with the fact that we may
identify $Y^a$ and $\tilde Y^a$ for a smooth vector field.
\enddemo

Suppose $\Con{a}{bc}$ is another connection, then covariant
differentiation with respect to it may be expressed as
$$ X^a{}_{;b}=X^a{}_{,b}+\Con{a}{bc} X^c $$
this may also be expressed as
$$ X^a{}_{;b}=X^a{}_{|b}+\hCon{a}{bc} X^c $$
where $\hCon{a}{bc}=\Con{a}{bc}-\con{a}{bc}$. The symbol $\hCon{a}{bc}$
will transform as a type $(1,2)$ tensor because it is the difference of two
connections.

If $X^a$, $Y^a$ and $\Con{a}{bc}$ are all smooth then $Y^bX^a{}_{|b}$
and $\hCon{a}{bc}X^cY^b$ commute with the embedding. This implies that
the covariant derivative $Y^bX^a{}_{;b}$ also commutes with the embedding.

We may define a torsion free connection $\hcon{a}{bc}$  by taking the
symmetric part of the background connection
$$ \eqalign{
   X^a{}_{:b} &= X^a{}_{|b} - \con{a}{[bc]} X^c \cr
   &= X^a{}_{,b} + \hcon{a}{bc} X^c \cr
}$$

For smooth tensor fields covariant differentiation with respect to the
torsion-free connection $\hcon{a}{bc}$ will commute with the embedding.
Rather than using the partial derivative to construct invariant objects, we
shall use the covariant derivative of the background torsion free connection$:a$.

The Lie bracket $[X,Y]$ of two vectors $X$ and $Y$ is independent of a
torsion free connection so one may write
$$ [X,Y]^a = X^b Y^a{}_{:b} - Y^b X^a{}_{:a} $$
since the covariant derivative $:a$ commutes with the embedding we have
$$ [\embed(X),\embed(Y)] = \embed([X,Y]) $$

As a second example we consider a smooth metric $g_{ab}$ and the
Levi-Civita connection
$$ \Con{a}{bc} = \tfrac12 g^{ad} ( g_{bd,c} + g_{cd,b} - g_{cd,b} ) $$
Using
$$ g_{ab:c} = g_{ab,c} - \hcon{d}{ac} g_{bd} - \hcon{d}{bc} g_{ad} $$
implies that
$$ \Con{a}{bc} = \hcon{a}{bc} + \tfrac12 g^{ad} ( g_{bd:c} + g_{cd:b} -
   g_{bc:d} )$$
therefore we may write
$$ X^a{}_{;b} = X^a{}_{:b} + \hCon{a}{bc} X^c $$
where
$$ \hCon{a}{bc} = \tfrac12 g_{ad} ( g_{bd:c}+ g_{cd:b} - g_{bc:d} )$$
Thus for a smooth metric the covariant derivative with respect to the
Levi-Civita connection commutes with the embedding.

\head
\secnum.
Association
\endhead
The relation of association or weak equivalence for generalised functions
is much the same as the normal definition; 

\definition{Definition \defnum}
We say that $[\tT]\in\CG\st{p}{q}(\Omega)$ is associated to zero (written as
$[\tT]\approx0$) if $\forall\Psi\in\tCD^q_p(\Omega)$, $\exists k>0$ such that
$$ \lim_{\eps\to0} \int {\tT\gt{a}{b}{b}{d}}(\Pxe,\gamma,x)
   \Psi\gt{c}{d}{a}{b}(x) \,dx =0 \qquad \forall 
   \Pex\in\CA0(\Real^n),\,\gamma\in\CK.$$
\enddefinition

Two generalised functions $[\tS]$ and $[\tT]$ are associated to each other
if $\tS-\tT\approx0$

\definition{Definition \defnum}
We say that $[\tT]\in\CG\st{p}{q}(\Omega)$ is associated to the distribution
$S\in\cD'\st{p}{q}(\Omega)$ (written as $[\tT]\approx S$) if
$\forall\Psi\in\tCD(\Omega)$, $\exists k>0$ such that
$$ \lim_{\eps\to0} \int {\tT\gt{a}{b}{b}{d}}(\Pxe,\gamma,x)
   \Psi\gt{c}{d}{a}{b}(x) \,dx = \tprod{S}{\Psi} \qquad \forall
   \Pex\in\CA0(\Real),\, \gamma\in\CK$$
\enddefinition

We finally verify that this assignment of a distributional interpretation
to a generalised function also commutes with $\mu$

\proclaim{Proposition \procnum}
$[\tT']\approx S'$ implies $[\muem\tT']\approx\mufun S'$
\endproclaim

\demo{Proof}
Letting $\Psi\in\tCD'{}^q_p(\Omega)$,
$$ \eqalign{
   & \limeps \int \muem \tT'\gt{a}{b}{c}{d}(\Phe,\gamma,x)
     \Psi\gt{c}{d}{a}{b} \, dx \cr
   & \qquad= \limeps \int \mux{a}{e}(x) \ldots \mux{b}{f}(x)
   \mmux{g}{c}(x) \ldots 
   \mmux{h}{d}(\eps) \tT'\gt{e}{f}{g}{h}(\muphi\Phe,\mu_*\gamma,\mu(x))
   \Psi\gt{c}{d}{a}{b}(x) \,dx \cr
   & \qquad= \limeps \int
   \tT'\gt{e}{f}{g}{h}(\muphi\Phe,\mu_*\gamma,x')
   \mu_*\Psi\gt{c}{d}{a}{b}(x) \,dx' \cr
   & \qquad= \tprod{S'}{\mu_*\Psi} \cr
   & \qquad= \tprod{\mufun S'}{\Psi} \cr
}$$
\enddemo

\head
\secnum.
Conclusion
\endhead
We now have a covariant theory of generalised functions in which we are
able to define tensors whose transformation laws coincide with those of
distributions, both in the senses of convolution embedding and
association.

This will mean that we may carry out calculations in General
Relativity that involve the evaluation of products of distributions, such
as the evaluation of the distributional curvature at the vertex of a cone
with a deficit angle $2\pi(1-A)$
by Clarke et al.\ (1996) in a coordinate independent way.
One would need to first choose a convenient coordinate system, then embed
the metric into the space $\CG\st02(\Real^2)$, which would transform
as a type $(0,2)$ tensor  using the definitions
in this paper, calculate the distributional curvature $[\tilde
R\sqrt{{\tilde g}}]$ as a generalised function, which will transform as a
scalar density of weight $+1$ and show that it is associated to the
distribution $4\pi(1-A)\delta^{(2)}$ obtained in that paper.

\head Acknowledgement \endhead
The authors wish to thank ESI for supporting their visit to the institute
as part of the project on \emph{Nonlinear Theory of Generalised functions}.
The authors thank M.~Kunzinger, M.~Oberguggenberger and R.~Steinbauer for
helpful discussions.  J.~Wilson acknowledges the support of EPSRC. grant
No. GR/82236.

\endcrossrefs
\enddocument

\bye